\documentstyle[12pt,plenum,epsf]{book}

\begin{document}
\chapter{HYDROGEN ATOMS IN NEUTRON STAR ATMOSPHERES:
ANALYTICAL APPROXIMATIONS FOR BINDING ENERGIES}

\author{Alexander Y. Potekhin\footnote{
e-mail: palex@astro.ioffe.rssi.ru}
}

\affiliation{Ioffe Physico-Technical Institute, 
St.-Petersburg 194021, Russia}

\section{INTRODUCTION}

Since the first observations of neutron stars
thirty years ago, they have affected many branches 
of physics. These extremely compact stars 
serve as natural physical laboratories
for probing the properties of matter under 
extreme physical conditions. 
In particular, more than half of them possess magnetic fields
$B>10^{12}$~G.

Despite their name, neutron stars consist 
not only of neutrons. They have a crust containing ionized iron,
heavier elements, and exotic neutron-rich nuclei,\refnote{\cite{Pethick}}
above which lie liquid and gaseous outer envelopes,
which are thought to be composed of iron or lighter 
elements.\refnote{\cite{CPY}}
The atmosphere, that affects the spectrum 
of outgoing thermal radiation, likely consists of 
hydrogen, the most abundant element in the Universe,
which might be brought to the star surface by fall-out of
circumstellar medium. Neutral atoms can provide an appreciable
contribution to the atmospheric opacity.

Apart from
the physics of neutron stars, 
quantum-mechanical calculations of strong\-ly
magnetized hydrogen atoms find application also in the physics
of white dwarf stars\refnote{\cite{WR87},\cite{Schweizer}} and in
the solid state physics.\refnote{\cite{Klaassen}}
Because of this practical demand, 
hydrogen in strong  magnetic fields
has been well studied
in the past two decades.\refnote{\cite{Ruder}}
The peculiarity of the problem for neutron stars
is that an atom cannot be considered abstractedly from 
its thermal motion. Indeed,  
neutron star atmospheres are hot ($T\sim10^5-10^6$~K),
so that typical kinetic energies of the atoms are non-negligible
in comparison with typical binding energies.
Taking the thermal motion into account is highly non-trivial,
because an atom moving across magnetic field 
is equivalent to an atom placed in orthogonal
electric and magnetic fields,
so that the cylindrical symmetry is broken.

At $\gamma\gg1$, where 
$\gamma\equiv\hbar\omega_c /2{\rm~Ryd}=B/2.35\times10^9\mbox{ G}\gg1$
and $\omega_c$ is the electron cyclotron frequency,
the collective motion 
effects\refnote{\cite{Avron},\cite{VB88}}
become especially pronounced. 
In particular, so-called decentered states 
(with the electron localized mostly in the ``magnetic well'' aside 
from the Coulomb center) are likely to be populated
even at the relatively high densities $\rho > 10^{-2}$ g cm$^{-3}$
typical of neutron star atmospheres.
These exotic states have been predicted two decades ago
by Burkova et al.\refnote{\cite{Burkova}}
and studied recently by other 
authors.\refnote{\cite{Dzyaloshinskii}--\cite{Schmelcher}}

Collective-motion effects on the usual ``centered'' states
have been first considered
in frames of the theory of perturbation.\refnote{\cite{VB88},\cite{PM93}}
Non-perturbative results covering both centered and 
decentered
states were subsequently presented
for binding energies and 
wavefunctions,\refnote{\cite{VDB92},\cite{P94}} 
oscillator strengths,\refnote{\cite{P94}}
spectral line shapes,\refnote{\cite{PP95}} and
photoionization cross sections.\refnote{\cite{PP97}}
None of these data, however, has been published 
in an easy-to-use form of tables or analytical expressions.
 
In this contribution I propose approximate analytical 
expressions for the 
binding energies of the hydrogen atom arbitrarily moving 
in a magnetic field typical of neutron stars, $300\leq\gamma\leq10^4$.
This range is physically distinguished, since at weaker fields
the spectrum is strongly complicated by multiple
narrow anticrossings,\refnote{\cite{VDB92}}
while the upper bound, $\gamma\sim10^4$,
corresponds to the onset of non-negligible relativistic 
effects.\refnote{\cite{Chen}}

\section{THEORETICAL FRAMEWORK}

Motion of the hydrogen atom in a magnetic field 
can be conveniently described by the pseudomomentum 
$ 
{\bf K} = {m_p} \dot{{\bf r}}_p + {m_e} \dot{{\bf r}}_e
 - ({e}/{c}) {\bf B} \times 
({\bf r}_e -{\bf r}_p),
$
where the subscript $i=e$ or $i=p$ indicates electron 
or proton, respectively, 
$ 
   \dot{{\bf r}}_i
   = -({\rm i}\hbar/m_i)\nabla_i 
   -(q_i/m_i c){\bf A}({\bf r}_i)
$
is the velocity operator, $m_i$ the mass, 
$q_p=-q_e=e$ the charge, and
${\bf A}({\bf r})$ the vector potential of the field. 
Gorkov and Dzyaloshinskii\refnote{\cite{GD68}} have shown that
in the representation in which all components of ${\bf K}$ 
have definite values, the relative motion 
can be described in terms of a one-particle Hamiltonian
which depends on ${\bf K}$. 

It is convenient to describe the centered states 
of the atom 
using the relative coordinate
${\bf r}^{(0)}={\bf r}_e-{\bf r}_p$ 
as independent variable and the 
axial gauge of the vector potential, 
${\bf A}({\bf r}) = \frac12{\bf B}\times {\bf r}$.
For the decentered states, the ``shifted''
representation\refnote{\cite{GD68}} is more convenient.
In the latter representation, 
the independent variable is 
${\bf r}^{(1)}={\bf r}_e-{\bf r}_p-{\bf r}_c$
and the gauge is
${\bf A}({\bf r}) = \frac12{\bf B}\times 
({\bf r}-\left[({m_p}-{m_e})/{m_{\rm H}}\right]{\bf r}_c)$. 
Here, 
${\bf r}_c= \frac{c}{eB^2}{\bf B}\times{\bf K}$
is the relative guiding center, and
${m_{\rm H}}={m_p}+{m_e}$.

Let us assume that ${\bf B}$ is directed 
along the $z$-axis. The $z$-component of the pseudomomentum 
corresponding to the motion
along the field yields the
familiar term $K_z^2/2{m_{\rm H}}$ in the energy,
while the transverse components ${\bf K}_\perp$
produce non-trivial effects.
Therefore we assume $K_z=0$ 
and ${\bf K}_\perp={\bf K}$ hereafter.

If there were no Coulomb attraction, then
the transverse part of the wavefunction 
could be described by a Landau function
$\Phi_{ns}({\bf r}^{(1)}_\perp)$, where
${\bf r}^{(1)}_\perp$ is the projection of ${\bf r}^{(1)}$
in the $(xy)$-plane.
The energy of the transverse excitation 
is
\begin{equation} 
   E^\perp_{ns} = [
   n+({m_e}/{m_p})(n+s)]
\hbar\omega_c,
\end{equation} 
where the zero-point and spin terms are disregarded. 

A wavefunction $\psi_\kappa$ of an atomic state 
$|\kappa\rangle$ can be expanded over the complete set of the 
Landau functions
\begin{equation} 
   \psi_\kappa^{(\eta)}({\bf r}^{(\eta)}) = 
   \sum_{ns} \Phi_{ns}({\bf r}_\perp^{(\eta)})\, 
   g^{(\eta)}_{n,s;\kappa}(z),
\label{expan}
\end{equation} 
where $\eta=0$ or 1 indicates the conventional or shifted representation,
respectively (a generalization to arbitrary $\eta$ 
proved to be less useful\refnote{\cite{P94}}).
The one-dimensional
functions $g^{(\eta)}_{ns;\kappa}$ are to be found
numerically. The adiabatic approximation used in early 
works\refnote{\cite{Burkova},\cite{GD68}}
corresponds to retaining
only one term in this expansion.

A bound state can be numbered\refnote{\cite{P94}} as 
$|\kappa\rangle = |n_\kappa,s_\kappa, \nu, {\bf K}\rangle$, 
where $n_\kappa$ and $s_\kappa$ relate to the leading term of the 
expansion (\ref{expan}), and $\nu$ enumerates longitudinal 
energy levels
\begin{equation} 
   E^\|_{n_\kappa,s_\kappa, \nu}(K) = E_\kappa - E^\perp_{n_\kappa s_\kappa}
\label{elong}
\end{equation} 
and controls the $z$-parity:
$g^{(\eta)}_{n,s;\kappa}(-z)=(-1)^\nu g^{(\eta)}_{n,s;\kappa}(z)$.
For the non-moving atom
at $\gamma>1$, the states $\nu=0$ are tightly bound 
in the Coulomb well, while the states $\nu\geq1$ 
are hydrogen-like, with binding energies below 1 Ryd.
The states with $n\neq0$ 
belong to continuum at $\gamma>0.2$ and will not
be considered here.

At small pseudomomenta $K$,
the states $\nu=0$ remain tightly bound and centered,
the mean electron-proton separation 
$\bar{x}$ being considerably smaller than $r_c$
(for the hydrogen-like states $\nu\geq 1$, however,
$\bar{x}$ is close to $r_c$ at any $K$).
The larger $K$, the greater is the distortion of 
the wavefunction towards ${\bf r}_c$,
caused by the motion-induced electric field
in the co-moving reference frame, until
near some $K_c$
transition to the decentered state occurs,
and the character of the motion totally changes.
With further increasing $K$,
the transverse velocity 
decreases and tends to zero, 
whereas the electron-proton separation 
increases and tends to $r_c$. 
Thus, for the decentered states, the pseudomomentum 
characterizes electron-proton separation
rather than velocity.

At very large $K$ the longitudinal functions
become oscillator-like, corresponding to
a wide, shallow parabolic potential well.\refnote{\cite{Burkova}}
For a fixed $\nu$, this limit is reached at 
$K\gg(\nu+\frac12)^2 \hbar/{a_{\rm B}}$, where ${a_{\rm B}}$ is the Bohr radius.
Still at arbitrarily large $K$, there remain infinite number
of bound states with high values of $\nu$
whose longitudinal wavefunctions are governed
by the Coulomb tail
of the effective one-dimensional potential.\refnote{\cite{P94}}

The decentered states of the atom 
at $K>K_c\sim10^2$~au have relatively low binding
energies and large quantum-mechanical sizes,
$l\sim K/\gamma$~au;
therefore they are expected to be destroyed 
by collisions with surrounding particles
in the laboratory and in the white-dwarf atmospheres.
In neutron-star atmospheres
at $\gamma\sim10^3$, however, the decentered states
may be significantly populated.
This necessitates inclusion of the entire
range of $K$ below and above $K_c$
in the consideration.

\section{ANALYTICAL APPROXIMATIONS}

\subsection{Binding Energies of the Non-Moving Hydrogen Atom}

Extensive tables of binding 
energies of the hydrogen atom at rest with respect
to the magnetic field 
have been presented by R\"osner et al.\refnote{\cite{Rosner}}
and supplemented by other 
authors.\refnote{\cite{Wintgen}--\cite{Xi}}
Recently, the accuracy $\sim10^{-12}$ Ryd
has been achieved.\refnote{\cite{Kravchenko}}
In the astrophysics,
a lower accuracy is usually sufficient,
and simple analytical estimates are often desirable. 

\atable{
Parameters of the approximation 
(\ref{e0gamma})
at $10^{-1}\leq\gamma\leq10^4$.\label{tab-e0}}
{\begin{tabular}{lllllllll}
\hline
$s$  & 0       & 1      & 2      & 3       & 4       
& 5       & 6       & 7        \\
\hline
$p_1$& 15.55   & 0.5332 & 0.1707 & 0.07924 & 0.04696 
& 0.03075 & 0.02142 & 0.01589 \\
$p_2$& 0.3780  & 2.100  & 4.150  & 6.110   & 7.640   
& 8.642   & 9.286   & 9.376   \\
$p_3$& 2.727   & 3.277  & 3.838  & 4.906   & 5.787   
& 6.669   & 7.421   & 8.087   \\
$p_4$& 0.3034  & 0.3092 & 0.2945 & 0.2748  & 0.2579  
& 0.2431  & 0.2312  & 0.2209  \\
$p_5$& 0.4380  & 0.3784 & 0.3472 & 0.3157  & 0.2977  
& 0.2843  & 0.2750  & 0.2682  \\
\hline
\end{tabular}
}

For this reason, we have constructed
a fit to $E^{(0)}$, 
where $E^{(0)}_{ns\nu}\equiv -E_{ns\nu}^\|(0)$, 
in a possibly widest range of $\gamma$.
For the tightly-bound states,
we have
\begin{equation} 
\hspace{-5em}
   E^{(0)}_{0s0}(\gamma)=\ln\left(
   \exp\left[(1+s)^{-2}\right]
   +p_1\left[\ln(1+p_2\sqrt{\gamma})\right]^2\right)
   +p_3\left[\ln(1+p_4\gamma^{p_5})\right]^2\mbox{~Ryd}.
\label{e0gamma}
\end{equation} 
The parameters $p_1-p_5$ depend on $s$; they are listed in
table \ref{tab-e0}. This fit is accurate to within 0.1--1\%
at $\gamma=10^{-1}-10^4$, and it also provides
the correct limits at $\gamma\to0$.

For the hydrogen-like states, 
we use the asymptotic result\refnote{\cite{Haines}}
\begin{equation} 
   E^{(0)}_{ns\nu}={\mbox{1 Ryd}\over (N+\delta)^2},
   \mbox{~where~}\left\{
   \begin{array}{ll}
   N=(\nu+1)/2, ~~~\delta\sim\gamma^{-1} & \mbox{for odd $\nu$,}\\
   N=\nu/2, ~~~~\delta\sim(\ln\gamma)^{-1}& \mbox{for even $\nu$}.
   \end{array}
   \right.
\label{e0h}
\end{equation} 
We have obtained the following fits to 
the quantum defect $\delta$: for odd $\nu$,
$ 
  \delta=
   (a_\nu+b_\nu\sqrt{\gamma}+0.077\gamma)^{-1},
$
where $a_\nu\approx1$ and $b_\nu\approx2$; and
for even $\nu$,
$ 
    \delta=
   [a_\nu+1.28\,\ln(1+b_\nu\gamma^{1/3})]^{-1},
$
where $a_\nu\approx\frac{2}{3}$ and $b_\nu\approx\frac{2}{3}$.
More accurate values of $a_\nu$ and $b_\nu$
are given in table \ref{tab-e0h}.
At $1\leq\gamma\leq10^4$, 
errors of these approximations
lie within $\sim10^{-3}$.

\subsection{Binding Energies of the Moving Hydrogen Atom}

For the moving hydrogen atom in a strong magnetic field,
the first analytical fit to $E(K)$ has been published by
Lai and Salpeter.\refnote{\cite{Lai}}
It is rather accurate for the ground state 
at $K<K_c$
but cannot be applied to excited or decentered states.

We describe the longitudinal 
energy (\ref{elong}) by the formula
\begin{equation} 
   |E^\|_{ns\nu}(K)| = 
   {E_{ns\nu}^{(1)}(K) \over 1+(K/K_c)^{1/\alpha}} 
   + {E_{ns\nu}^{(2)}(K) \over 1+(K_c/K)^{1/\alpha}}.
\label{eappr0}
\end{equation} 
The two-term structure of (\ref{eappr0}) is dictated by
the necessity to describe the two physically distinct
regions of $K$ below and above $K_c$.
The parameter $\alpha$ has the meaning of the width 
of the transition region near $K_c$
in logarithmic scale of pseudomomenta. 

\atable{
Parameters of
(\ref{e0h}) at $1\leq\gamma\leq10^4$.\label{tab-e0h}}
{\begin{tabular}{lllllll}
\hline
$\nu$          & 1     & 2     & 3     & 4     & 5     & 6     \\
\hline
$a_\nu$        & 0.785 & 0.578 & 0.901 & 0.631 & 0.970 & 0.660 \\
$b_\nu$        & 1.724 & 0.765 & 1.847 & 0.717 & 1.866 & 0.693 \\
\hline
\end{tabular}
}

For the tightly-bound states, we parameterize 
the dependencies $E^{(j)}(K)$ as follows:
\begin{equation}
   E_{0s0}^{(1)}(K) = E^{(0)}_{0s0}-
   {K^2\over 2m_{\rm eff}+q_1 K^2/E^{(0)}_{0s0} },
~~~~
   E_{0s0}^{(2)}(K) =
   {2{\rm~Ryd}\over \sqrt{r_\ast^2+r_\ast^{3/2}+q_2 r_\ast}},
\label{eappr12}
\end{equation}
where $r_\ast =r_c/{a_{\rm B}}=K/(\gamma$~au),
$q_1$ and $q_2$ are dimensionless fitting parameters,
and $m_{\rm eff}$ is the effective mass
which is close to (but not necessarily coincident with)
the transverse effective mass $M^\perp_{ns\nu}$
obtained by the perturbation technique.
At $\gamma\geq300$,
we put
$q_1=\log_{10}(\gamma/300)$ if $s=0$ and
$q_1=0.5$ otherwise,
$q_2=0.158\,[\ln((1+0.1s)\gamma/215)]^{2/5}$, and
$\alpha=0.053\,\ln(\gamma/150)$.
For the effective mass, we have
$ 
   m_{\rm eff}={m_{\rm H}}\left[1+(\gamma/\gamma_0)^{c_0}\right],
$
where $c_0=0.937+0.038s^{1.58}$ and 
$\gamma_0 = 6150(1+0.0389s^{3/2})/
          [1+7.87s^{3/2}]$.
For the critical pseudomomentum, 
we have
$K_c=[c_1+\ln(1+\gamma/\gamma_1)]\sqrt{2{m_{\rm H}} E^{(0)}}$.
The parameters $c_1$ and $\gamma_1$ take on the values
$c_1=0.81,1.09,1.18,1.24$
and
$\gamma_1=(8.0,3.25,2.22,1.25)\times10^4$
for
$s=0,1,2,3$, respectively.
For $s\geq4$, we put $c_1=0.93+0.08s$ and $\gamma_1=6500$.

\begin{figure}
\begin{center}
\epsfysize=75mm
\epsfbox[50 310 330 560]{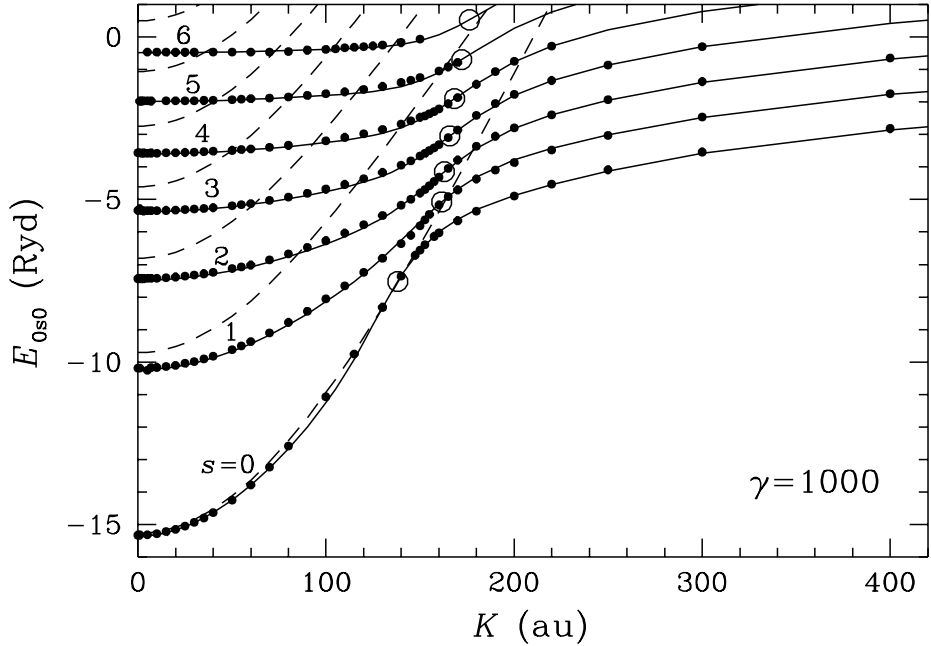}
\end{center}
%%%\vspace*{90mm}
\caption{
Energy spectrum of the hydrogen atom moving across
strong magnetic fields.
Numerical values (dots) are compared with 
the present analytical approximations 
(full lines) and with previously published\refnote{\cite{Lai}} ones
(dashed lines).
\label{fig1}
}
\end{figure}

In figure~\ref{fig1} the above fitting formulae are compared 
with our numerical results\refnote{\cite{P94}}
and with the previous approximations.\refnote{\cite{Lai}} 
The figure demonstrates that the present approximations
are valid at any $K$ from 0 to infinity.
Appreciable discrepancies 
occur only in narrow ranges of $K$ near anticrossings.

Now let us turn to the hydrogen-like states.
Their binding energies
are approximated by the same formula (\ref{eappr0})
but with slightly different expressions
for $E^{(1)}$ and $E^{(2)}$. For these states,
$M^\perp_{ns\nu}$ exceeds ${m_{\rm H}}$ by orders of magnitude,
and the perturbation method fails already 
at small $K$,\refnote{\cite{PM93}} 
rendering the notion of the effective mass 
practically useless for the fitting. 
Thus we consider $m_{\rm eff}$ as effectively infinite
and put $E^{(1)}_{0s\nu}(K)=E^{(0)}_{0s\nu}$ ($\nu\geq1$).
Furthermore, the transition region is not well defined,
and therefore $K_c$ and $\alpha$ lose
their clear meaning and become mere
fitting parameters. For odd states, we have, approximately,
$K_c=(\nu^{5/4}\gamma/170)^{0.9}\sqrt{2{m_{\rm H}} E^{(0)}}$ and 
$\alpha=0.66+\nu/20$. For even states,
$K_c=\nu\,\sqrt{(\gamma/600){m_{\rm H}} E^{(0)}}$ and
$\alpha=0.66$.

The function $E^{(2)}(K)$ that describes the longitudinal energy
at large $K$ is now
\begin{equation} 
      E_{0s\nu}^{(2)}(K) =
   \left\{(2{\rm~Ryd})^{-1}\left[
   r_\ast^2+(2\nu+1)r_\ast^{3/2}+q_2 r_\ast 
   \right]^{1/2}
   +1/E^{(0)}_{0s\nu}\right\}^{-1},
\label{eappr3}
\end{equation} 
with
$q_2=\nu^2-1$ for odd $\nu$, and
$q_2=\nu^2+2^{\nu/2}\log_{10}(\gamma/300)$ for even $\nu$ 
(at $\gamma\geq300$).
The first and second terms in the square brackets ensure
the correct asymptotic behavior.\refnote{\cite{P94}}

\section{CONCLUDING REMARKS}

The analytical approximations for binding energies
presented in this contribution
depend continuously on two arguments ---
magnetic field strength and transverse pseudomomentum.
They are accurate, typically, within a few parts in 100--1000.
The accuracy can be improved by almost an order of magnitude
by optimizing the parameters $m_{\rm eff}$, $K_c$, $\alpha$,
$q_1$, $q_2$  
in equations 
(\ref{eappr0})--(\ref{eappr3})
separately at each discrete value of $\gamma$. 
Tables of such optimized
parameters have been obtained and will be published elsewhere,
together with analytical approximations of 
geometrical sizes of various quantum-mechanical states of the moving
atom and oscillator strengths of radiative transitions
among them. The atomic sizes play important role
in distribution of atoms over quantum states in a plasma
and in their contribution to the plasma absorption 
coefficients. For example, a size of an atom
may be used to evaluate effects of ``unbounding''
of electrons caused by random charge distribution in the plasma.
For non-magnetized hydrogen plasma, an approximate treatment of
these effects was revised recently;\refnote{\cite{P96}}
for the strong magnetic fields analogous work is under way.
Eventually, the analytical estimates of $K$-dependencies of the 
binding energies, atomic sizes, and transition rates
help to generalize previously developed models of
fully ionized atmospheres of magnetic neutron 
stars\refnote{\cite{Shibanov}}
to the more realistic case of partially ionized atmospheres.

{\referencestyle

}
\end{document}